\documentclass[pra,twocolumn,amsmath,amssymb,superscriptaddress]{revtex4-1}

\usepackage{epsfig,amsmath}
\usepackage{subfigure}
\usepackage{graphicx}
\usepackage{dcolumn}
\usepackage{stmaryrd}
\usepackage{mathrsfs}
\usepackage{pifont}
\usepackage{amsthm}
\usepackage{amssymb}
\usepackage{bm}
\usepackage{latexsym}
\usepackage[colorlinks=true,linkcolor=blue,citecolor=blue]{hyperref}
\usepackage{color}
\usepackage{epstopdf}

\theoremstyle{plain}

\newcommand{\Lam}{\Lambda}
\newcommand{\de}{\delta}
\newcommand{\De}{\Delta}

\newcommand{\non}{\nonumber}
\newcommand{\ti}{\tilde}
\newcommand{\la}{\langle}
\newcommand{\ra}{\rangle}

\newcommand{\Ga}{\Gamma}
\newcommand{\om}{\omega}

\newcommand{\si}{\sigma}

\newcommand{\al}{\alpha}
\newcommand{\be}{\beta}
\newcommand{\ep}{\epsilon}

\newcommand{\mL}{\mathcal{L}}

\newcommand{\mU}{\mathcal{U}}

\newcommand{\mM}{\mathcal{M}}

\begin{document}

\title{Measurement-induced cooling of a qubit in structured environments}

\author{Jia-Ming Zhang}
\affiliation{Zhejiang Province Key Laboratory of Quantum Technology and Device, Department of Physics, Zhejiang University, Hangzhou 310027, Zhejiang, China}

\author{Jun Jing}
\email{jingjun@zju.edu.cn}
\affiliation{Zhejiang Province Key Laboratory of Quantum Technology and Device, Department of Physics, Zhejiang University, Hangzhou 310027, Zhejiang, China}

\author{Lian-Ao Wu}
\affiliation{Department of Theoretical Physics and History of Science, The Basque Country University (UPV/EHU), P.O. Box 644, 48080 Bilbao, Spain}
\affiliation{Ikerbasque, Basque Foundation for Science, 48011 Bilbao, Spain}

\author{Li-Gang Wang}
\affiliation{Zhejiang Province Key Laboratory of Quantum Technology and Device, Department of Physics, Zhejiang University, Hangzhou 310027, Zhejiang, China}

\author{Shi-Yao Zhu}
\affiliation{Zhejiang Province Key Laboratory of Quantum Technology and Device, Department of Physics, Zhejiang University, Hangzhou 310027, Zhejiang, China}

\date{\today}

\begin{abstract}
In this work, we study the dynamics of a two-level system (qubit) embedded in a finite-temperature structured bath under periodical nondemolition measurements. The system under measurements will reach a quasisteady state, whose effective temperature can be maintained lower than that of the surrounding environment. We study the influence of the environmental modes from different regimes of frequency on the qubit. The spectrum of the bath consisting of a large number of bosonic harmonic oscillators can be approximately divided into three parts according to their effects of cooling or heating. Due to the spectral analysis over the structured bath based on a time-convolutionless master equation beyond the rotating-wave approximation, we propose a necessary cooling condition for the bath in the context of quantum nonselective and nondemolition measurements. It consists of two parts: (i) the logarithmic derivative of the spectrum around the system transition frequency should be large enough, at least larger than one-half of the inverse temperature of the bath; (ii) the spectrum should have a sharp high-frequency cutoff that cannot be far detuning from the system transition frequency. From this condition, we find that environments with two popular types of spectra, i.e., the modified Lorentzian models and the super-Ohmic models, are available for cooling the open quantum system.
\end{abstract}

\maketitle

\section{Introduction}\label{intro}

The unavoidable coupling between the open quantum system and the environment allows the environment to thermalize the quantum system~\cite{Breuer2002,Gardiner2004,Scully2012}. Cooling the quantum system has long been a challenge and one of the most desirable quantum technologies. It plays a crucial role in the initialization of the quantum applications including but not limited to the adiabatic quantum computing~\cite{Childs2001,Farhi2001,Sarandy2005,Hammerer2009,You2011} and ultrahigh-precision measurements using mechanical resonators~\cite{Bocko1996,Caves1980,Li2011}.

A passive and straightforward way of cooling a quantum system is attaching the system to a cold bath. To cool down a system to an even lower temperature, a successful approach called sideband cooling employs transitions from the upper state of the system, e.g., a qubit, to an auxiliary intermediate state, which can then quickly relax into the ground state~\cite{Chu1998,Cohen-Tannoudji1998,Phillips1998,Leibfried2003,Valenzuela2006,Wineland1978,Neuhauser1978,Monroe1995}. One disadvantage of this cooling method is the requirement of an appropriate intermediate state. An alternative is sponsored by selective quantum measurements~\cite{Lloyd1997,Scully2001,Li2011,Xu2014,Pyshkin2016}. The results of measurement are read out to discard the unwished samples. One major disadvantage of this approach is that it requires performing many measurements to achieve the cooling target, and the survival probability is quite small.

Since the quantum systems cannot be isolated from the surrounding environment, an interesting idea is to exploit the environment to cool them down. In the spirit of this idea, frequent quantum measurements modifying the thermodynamic properties of the environment have been proposed theoretically~\cite{Erez2008,Gordon2009,Gordon2010,Gelbwaser2014,Kurizki2015,Kurizki2017} and verified experimentally~\cite{Gonzalo2010}. In the open-quantum-system theory, the Markovian limit renders an invariant and unidirectional energy-flow rate from the system to the environment during the process that the system is cooled down to the thermal-equilibrium state. However, there are more numbers of scenarios where the relaxation time scale of the environment is not sufficiently small compared to the dynamical time scale of the system, which means that the bath would have a memory effect on the system. The environmental memory effect has been used to extract work from the bath~\cite{Gelbwaser2013}, exceed the classical Carnot bound~\cite{Zhang2014}, and freeze the system state~\cite{Gordon2010}.

On account of the memory capacity of the structured baths, the energy flow may go from the system into the environment and then come back to the system until reaching the thermal equilibrium. More importantly, the structured baths are available to perform manipulation over the dynamics of the open system. The free unitary evolution of the combined density matrix $\rho$ of the system and the environment over a time $\tau$ can be expressed by $\mU_{\tau}[\rho(0)]$. Then supposing that the system and the environment could be roughly decoupled at this moment by, e.g., a nonselective impulsive measurement in the basis of the system bare Hamiltonian, the combined density matrix now becomes $\mM\mU_{\tau}[\rho(0)]$. Assume the time interval $\tau$ is so properly chosen that at the end of this interval, the energy flowing from the system to the environment exceeds that moving along the reverse direction, then an amount of net energy of the system would be retained in the environment by the measurement. Performing a sequence of periodical measurements with a constant separation time $\tau$ while among the measurements the system freely evolves under the environment; then the combined density matrix can be written as
\begin{equation}\label{process}
\rho(t=n\tau)=(\mM\mU_{\tau})^n[\rho(0)]
\end{equation}
after $n$ consecutive measurements. The process that system energy flows into the environment repeated by periodic quantum measurements pushes the quantum system into a quasisteady state. The effective temperature of the system is controllable by the measurement frequency and in certain conditions becomes even lower than the environmental temperature. This method may be realized in many existing experimental scenarios, such as microcavities and quantum dots~\cite{Gordon2009}. The cooling phenomenon was attributed to the quantum anti-Zeno effect~\cite{Erez2008}, yet we will see this argument is controversial.

Recently, we proposed a compact criterion concerning the spectral density function (SDF) of a zero-temperature environment to discriminate quantum Zeno and anti-Zeno effects~\cite{Zhang2018}. Inspired by a similar spectral analysis, we find in this work that the bath structure obviously affects the availability and efficiency of system cooling, and the cooling effect is not equivalent to the anti-Zeno effect. In a finite-temperature environment, we find that the cooling effect arising for a short measurement interval entails the breakdown of the rotating-wave approximation (RWA) and nontrivial contribution from the time-dependent damping rate of the system. The counter-rotating terms impact the final quasisteady state and could not be neglected. We examine the spectrum of the environment and conclude two conditions contributing to cooling: (i) the logarithmic derivative of the spectrum at the system transition frequency is large enough; (ii) the spectrum should have a sharp high-frequency cutoff.

The rest of this work is organized as follows. In Sec.~\ref{dyna}, we focus on the free evolution $\mU_{\tau}(\rho)$ described by a time-convolutionless master equation, which briefly introduces the dynamics of the system in a finite-temperature bath without RWA. We analyze the dynamical contributions from the rotating-wave and counter-rotating terms in details. Section~\ref{thdy} is devoted to the system dynamics under periodic nondemolition measurements. The connection and distinction between the quantum anti-Zeno effect and cooling are clarified. In Sec.~\ref{cool}, we establish the cooling conditions by spectral analysis and whereby study the cooling phenomenon in the modified Lorentzian model and the super-Ohmic model. We close this work with a summary in Sec.~\ref{conc}.

\section{Evolution of the system without measurements}\label{dyna}

\subsection{Time-convolutionless master equation for the system in a finite-temperature bath without RWA}

We consider a two-level system (qubit or TLS) with Bohr frequency $\om_a$ undergoing decay into a finite-temperature bath. The bath can be represented by a set of bosonic harmonic oscillators. The total Hamiltonian in the Schr\"{o}dinger picture has a general form ($\hbar=1$),
\begin{eqnarray}\non
H&=&\frac{1}{2}\om_a\si_z +\sum_k\om_k a_k^\dag a_k+\sum_k(g_k \si_+ a_k + g_k^*\si_- a_k^\dag) \\ \label{H}
&+&\sum_k(g_k \si_- a_k + g_k^* \si_+ a_k^\dag),
\end{eqnarray}
where $\si_z$ and $\si_{\pm}$ are respectively the Pauli matrix and the inversion operators of the system, $a_k^\dag$ and $a_k$ are respectively the creation and annihilation operators for the $k$th mode of the environment with frequency $\omega_k$, and $g_k$ describes the coupling strength between the system and the $k$th mode. The interaction Hamiltonian contains both the rotating-wave terms $\si_+a_k$ and $\si_-a_k^\dag$ and the counter-rotating terms $\si_+a_k^\dag$ and $\si_-a_k$.

In the weak-coupling regime, the evolution of the reduced density matrix of the system $\rho_S(t)$ over time $t$ can be written in the Schr\"{o}dinger picture as (a detailed and general derivation is presented in Appendix~\ref{SNME})
\begin{equation}
\label{ME}
\begin{aligned}
\frac{d}{dt}\rho_S(t)=&-i\left[\frac{1}{2}\om_a\si_z +\sum_{j=\pm}\De_j(t)\si_j^\dag \si_j,\rho_S(t)\right]\\
&+\sum_{j=\pm}
\frac{\Ga_j(t)}{2}\mL[\si_j](\rho_S(t))\\
&+\sum_{j=\pm}\left\{\left[\frac{\Ga_j(t)}{2}+i\De_j(t)\right]\si_j\rho_S(t)\si_j
+{\rm H.c.}\right\}.
\end{aligned}
\end{equation}
Here $\De_{+(-)}(t)$ and $\Ga_{+(-)}(t)$ are respectively the time-dependent Lamb shift of the ground (existed) state $|g\ra$ ($|e\ra$) and the time-dependent transition rate for $|g\ra\to|e\ra$ ($|e\ra\to|g\ra$). All of these coefficients can be decomposed into two parts: $\De_{\pm}(t)=\De^r_{\pm}(t)+\De^{cr}_{\pm}(t)$ and $\Ga_{\pm}(t)=\Ga^r_{\pm}(t)+\Ga^{cr}_{\pm}(t)$. Note, throughout this work, the superscripts $r$ and $cr$ are respectively used to signify the contributions from the rotating-wave terms and the counter-rotating terms in the interaction Hamiltonian and the subscripts $+$ and $-$ are respectively used to signify the contributions from the transitions $|g\ra\to|e\ra$ and $|e\ra\to|g\ra$. The Lamb shifts and the transition rates are respectively defined as
\begin{equation}
\label{De}
\begin{aligned}
\De_+^r(t)\equiv&\int_0^\infty d\om n_T(\om)G_0(\om)\frac{1-\cos[(\om-\om_a)t]}{\om-\om_a},\\
\De_+^{cr}(t)\equiv&-\int_0^\infty d\om [n_T(\om)+1]G_0(\om)\frac{1-\cos[(\om+\om_a)t]}{\om+\om_a},\\
\De_-^r(t)\equiv&-\int_0^\infty d\om [n_T(\om)+1]G_0(\om)\frac{1-\cos[(\om-\om_a)t]}{\om-\om_a},\\
\De_-^{cr}(t)\equiv&\int_0^\infty d\om n_T(\om)G_0(\om)\frac{1-\cos[(\om+\om_a)t]}{\om+\om_a},
\end{aligned}
\end{equation}
and
\begin{equation}
\label{Ga}
\begin{aligned}
\Ga_+^r(t)\equiv&2t\int_0^\infty d\om n_T(\om)G_0(\om){\rm sinc}[(\om-\om_a)t],\\
\Ga_+^{cr}(t)\equiv&2t\int_0^\infty d\om [n_T(\om)+1]G_0(\om){\rm sinc}[(\om+\om_a)t],\\
\Ga_-^r(t)\equiv&2t\int_0^\infty d\om [n_T(\om)+1]G_0(\om){\rm sinc}[(\om-\om_a)t],\\
\Ga_-^{cr}(t)\equiv&2t\int_0^\infty d\om n_T(\om)G_0(\om){\rm sinc}[(\om+\om_a)t],
\end{aligned}
\end{equation}
where $n_T(\om)=(e^{\be\om}-1)^{-1}$ is the temperature-dependent ($\be=1/T$ with $k_B\equiv1$) average population of the oscillator (bath mode) with frequency $\om$ at temperature $T$, and $G_0(\om)=\sum_k|g_k|^2\de(\om-\om_k)$ is the SDF at zero temperature. The Lindblad superoperators in the second line of Eq.~(\ref{ME}) are defined as
\begin{equation}
\mL[\si](\rho)\equiv 2\si\rho\si^\dag-\{\si^\dag \si,\rho\},
\end{equation}
representing the quantum jump process of the TLS characterized by an arbitrary system operator $\si$. The last line of Eq.~(\ref{ME}) represents the contribution of the so-called nonsecular terms and involves the two-photon processes. Note this line also includes $\si_+\rho_S(t)\si_+$ and $\si_-\rho_S(t)\si_-$, which stem from the cross interaction between the rotating-wave and the counter-rotating terms.

We can then obtain the generalized Bloch equations for the elements in $\rho_S(t)$ straightforwardly from Eq.~(\ref{ME}). The diagonal terms evolve according to~\cite{Kofman2004}
\begin{equation}\label{re}
\frac{d}{dt}\rho_{ee}(t)=-\frac{d}{dt}\rho_{gg}(t)=-\Ga_-(t)\rho_{ee}(t)+\Ga_+(t)\rho_{gg}(t),
\end{equation}
whereas the off-diagonal terms evolve according to
\begin{equation}\label{reg}
\begin{aligned}
&\frac{d}{dt}\rho_{eg}(t)=\frac{d}{dt}\rho_{ge}^*(t)\\
=&-\left\{\frac{\Ga_-(t)+\Ga_+(t)}{2}+i[\om_a+\De_-(t)-\De_+(t)]\right\}\rho_{eg}(t)\\
&+\left\{\frac{\Ga_-(t)+\Ga_+(t)}{2}-i[\De_-(t)-\De_+(t)]\right\}\rho_{ge}(t).
\end{aligned}
\end{equation}
It is clear that the time-dependent Lamb shift $\De(t)$ involves only the dephasing process of the system and the time-dependent transition rate $\Ga(t)$ affects both population decay and dephasing processes. The cooling or heating issue in this work resolves around Eq.~(\ref{re}) for populations and the retaining of the time dependence of the damping rates $\Ga_{\pm}(t)$ is crucial to cooling.

\begin{figure}[htbp]
\centering
\includegraphics[width=0.5\textwidth]{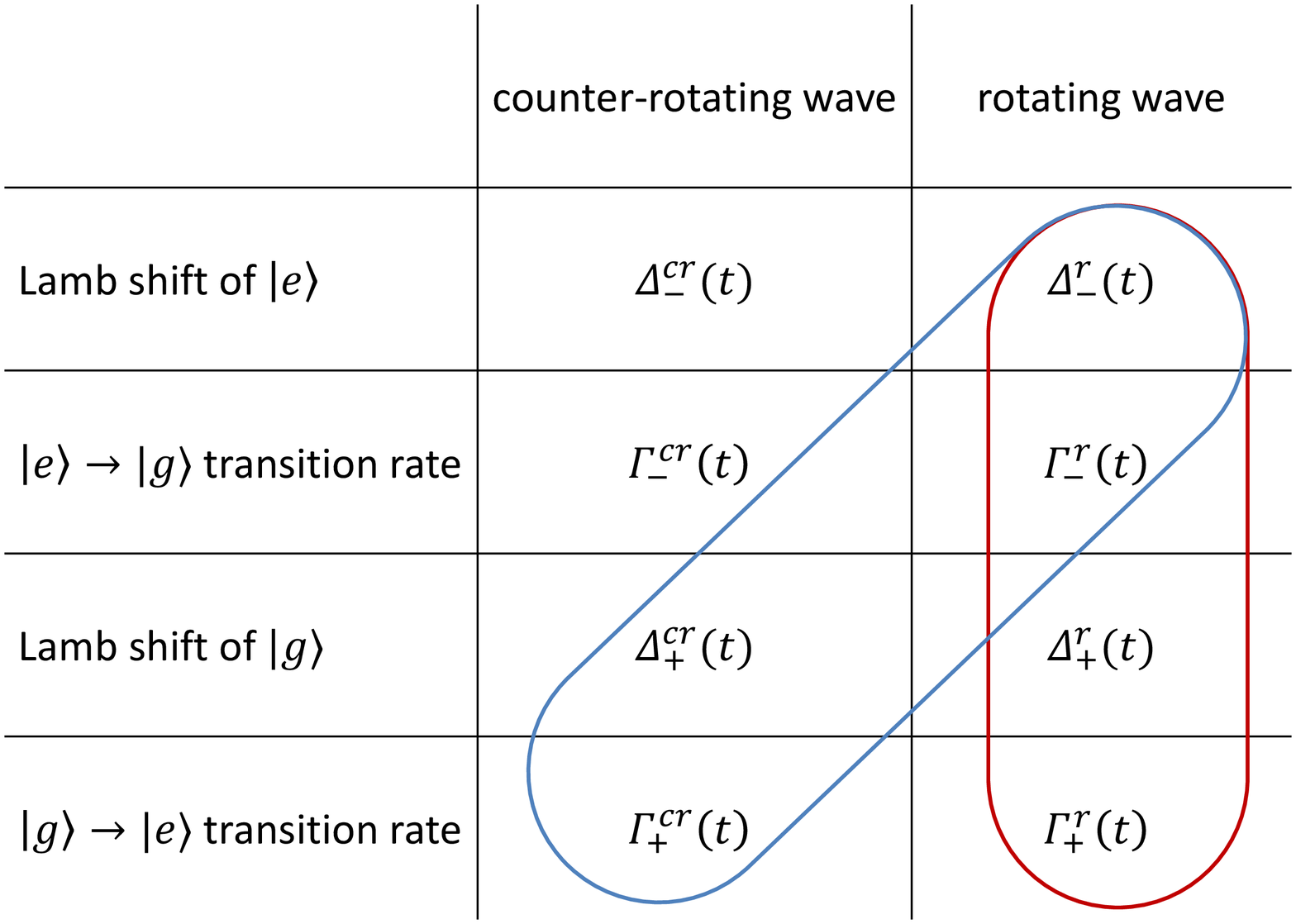}
\caption{(Color online) Transition rates and Lamb shifts appearing in the master equation~(\ref{ME}). They are relevant to various physics. The time-convolutionless master equation includes all of the eight items. With the rotating-wave approximation, the items with superscript $r$ are sustained, which are distinguished by the red rounded rectangle, but the remaining four items as well as those in the third line of Eq.~(\ref{ME}) vanish. For the zero-temperature environment or the vacuum-field bath, $\De_-^r(t)$, $\Ga_-^r(t)$, $\De_+^{cr}(t)$, and $\Ga_+^{cr}(t)$ are sustained, which are distinguished by the blue rounded rectangle, but the remaining four items vanish. With the vacuum field under RWA, only two items $\De_-^r(t)$ and $\Ga_-^r(t)$ will survive, which are the overlap of the red and blue rounded rectangles. All of these coefficients will become time independent in the Markovian limit. }
\label{NME}
\end{figure}

In different scenarios, the perturbative and time-convolutionless master equation~(\ref{ME}) can be reduced to simpler forms as we summarize in the sketch of Fig.~\ref{NME}. The last and nonsecular term of Eq.~(\ref{ME}) will be ignored by the secular approximation and then our master equation reduces to a master equation in the Lindblad form. The counter-rotating terms in Hamiltonian~(\ref{H}) will be canceled by RWA. Consequently, counter-rotating contributions in both Lamb shift $\De^{cr}(t)$ and state-transition rate $\Ga^{cr}(t)$ will disappear from Eq.~(\ref{ME}), as well as the terms for the cross interaction. For a zero-temperature environment (vacuum-field bath), i.e., in the limit $\be\to\infty$ and $n_T(\om)\to 0$, we have $\De_+^r(t)=\De_-^{cr}(t)=\Ga_+^r(t)=\Ga_-^{cr}(t)=0$. Physically, the vacuum-field assumption gives rise to partial loss of both transition rate for $|g\ra\to|e\ra$ and energy shift for $|g\ra$ due to the rotating-wave terms. It also causes partial loss of both transition rate for $|e\ra\to|g\ra$ and energy shift for $|e\ra$ due to the counter-rotating terms. With the Markovian approximation, which is valid for a long time scale and popular in literature regarding the effect of external environment, all of the time-independent energy shifts and transition rates in the master equation~(\ref{ME}) will become time independent.

\subsection{Thermal contributions}\label{thermal}

In the general framework of open-quantum-system dynamics, the environment is supposed to be at the thermal equilibrium state with temperature $T$. The bath can be described by a canonical ensemble and the constituent oscillators satisfy the Bose-Einstein statistics. Here we assume that the ground- and excited- state populations of the qubit similarly obey the canonical distribution to define its time-dependent effective temperature $T_S(t)$. The excited population thus satisfies the Fermi-Dirac statistics
\begin{equation}
\rho_{ee}(t)=\frac{1}{e^{\be_S(t)\om_a}+1},
\end{equation}
where $\be_S(t)\equiv1/T_S(t)$. In the following, we will use the time evolution of the excited population $\rho_{ee}(t)$ to characterize that of the effective temperature. The heating and cooling of the qubit correspond to the enhancement and declination of the excitation population $\rho_{ee}(t)$, respectively.

In a very short times cale, one can apply an adiabatic approximation to Eq.~(\ref{re}), such that $\rho_{ee}(t)\simeq\rho_{ee}(0)$. The solution of the excitation population $\rho_{ee}(t)$ in Eq.~(\ref{re}) then can be written as
\begin{equation}\label{ree}
\rho_{ee}(t)\approx\rho_{ee}(0)\left[1+e^{\be_S(0)\om_a}J_+(t)-J_-(t)\right],
\end{equation}
where $J_-(t)=J_-^r(t)+J_-^{cr}(t)$ and $J_+(t)=J_+^r(t)+J_+^{cr}(t)$ with
\begin{equation}\label{J}
\begin{aligned}
J_+^r(t)\equiv&\int_0^{t} dt' \Ga_+^r(t')\\
=&t^2\int_0^\infty d\om n_T(\om)G_0(\om){\rm sinc}^2\left[\frac{(\om-\om_a )}{2}t\right],\\
J_+^{cr}(t)\equiv&\int_0^{t} dt' \Ga_+^{cr}(t')\\
=&t^2\int_0^\infty d\om [n_T(\om)+1]G_0(\om){\rm sinc}^2\left[\frac{(\om+\om_a )}{2}t\right],\\
J_-^r(t)\equiv&\int_0^{t} dt' \Ga_-^r(t')\\
=&t^2\int_0^\infty d\om [n_T(\om)+1]G_0(\om){\rm sinc}^2\left[\frac{(\om-\om_a )}{2}t\right],\\
J_-^{cr}(t)\equiv&\int_0^{t} dt' \Ga_-^{cr}(t')\\
=&t^2\int_0^\infty d\om n_T(\om)G_0(\om){\rm sinc}^2\left[\frac{(\om+\om_a )}{2}t\right].
\end{aligned}
\end{equation}
Thus, for various initial population, one can find quite similar dynamics for $\rho_{ee}(t)$ (see the inset of Fig.~\ref{rL}). Note that even if initially the temperature of the system is equal to that of the environment, the state of the system still evolves with time for it will become entangled with the environment before they approach a new thermal equilibrium. The excitation population of the system remains unchanged under the Born-Markovian approximation, yet it will change with time under a structured environment as displayed by the black solid line in the inset of Fig.~\ref{rL}.

The relative deviation of excitation population at the moment $t$ is
\begin{equation}\label{dr}
\frac{\rho_{ee}(t)-\rho_{ee}(0)}{\rho_{ee}(0)}=\int_0^\infty d\om F[\be_S(0),\be,t,\om]G_0(\om),
\end{equation}
where the filter function $F$ could also be divided into the rotating-wave and counter-rotating components $F(\be_S,\be,t,\om)=F^r(\be_S,\be,t,\om)+F^{cr}(\be_S,\be,t,\om)$. They are respectively defined as
\begin{equation}\label{filter}
\begin{aligned}
F^r(\be_S,\be,t,\om)\equiv&t^2\frac{e^{\be\om}-e^{\be_S\om_a}}{e^{\be\om}-1}{\rm sinc}^2\left(\frac{\om-\om_a}{2}t\right),\\
F^{cr}(\be_S,\be,t,\om)\equiv&t^2 e^{\be_S\om_a}\frac{e^{\be\om}-e^{-\be_S\om_a}}{e^{\be\om}-1}{\rm sinc}^2\left(\frac{\om+\om_a}{2}t\right).
\end{aligned}
\end{equation}
The contribution of the rotating-wave terms representing the energy exchange between the qubit and the environment can be further divided into two parts with respect to the bath-mode frequency $\om$. The oscillators with frequency higher than $\om_a\be_S(0)/\be$ (it is assumed to be a boundary separating the low- and high-frequency domains) render a negative $F^r[\be_S(0),\be,t,\om]$, that can be used to cool down the qubit, whereas the oscillators with frequency lower than $\om_a\be_S(0)/\be$ render a positive $F^r[\be_S(0),\be,t,\om]$, meaning heating up the qubit. In the mean time, the fluctuation of the thermal field, which comes from the counter-rotating terms or the so-called energy nonconserving terms in the interaction Hamiltonian, generates virtual particles. In the case without initial population inversion of the system, that is to say, in the case of non-negative temperature, the function $F^{cr}[\be_S(0),\be,t,\om]$ is found to be always positive implying that the virtual process carries energy to the qubit and then heats it up.

Especially when $\be_S(0)=\be$, the boundary separating the positive and negative regimes of $F^r(\be,\be,t,\om)$ is just the system transition frequency $\om_a$. Combining the contributions from the rotating-wave and the counter-rotating filter functions in the high-frequency ($\om>\om_a$) regime, the upper bound for the cooling is found to be determined by the condition when $F^r(\be,\be,\tau,\om)+F^{cr}(\be,\be,\tau,\om)\leq0$. In the low-temperature limit, it can be estimated by
\begin{equation}\label{wup}
\om\lesssim\om_a[1+4\be\om_a\exp(-\be\om_a)].
\end{equation}
which is obtained using the mean value $1/2$ to replace the square sine function in Eq.~(\ref{filter}). Thus the bath spectrum can be roughly divided into three parts according to the cooling and heating effects. Low- and high-frequency oscillators with $\om<\om_a$ and $\om>\om_a[1+4\be\om_a\exp(-\be\om_a)]$ effectively heat up the qubit and those with moderate frequencies $\om_a<\om<\om_a[1+4\be\om_a\exp(-\be\om_a)]$ turn out to cool down the qubit.

It is noted that Eq.~(\ref{ree}) is obtained under the adiabatic approximation, so that the above interpretation is appropriate for a short time scale. Scrutinizing the filter functions $F^r(\be_S,\be,t,\om)$ and $F^{cr}(\be_S,\be,t,\om)$, the factor $e^{\be_S\om_a}$ in $F^{cr}(\be_S,\be,t,\om)$ indicates that the contribution from the counter-rotating terms overwhelms that from the rotating-wave terms especially at an extremely low temperature. Thus one should be very cautious when applying the rotating-wave approximation with short timescales and finite temperatures.

\section{Excitation dynamics under non-demolition measurements}\label{thdy}

In this section we consider the complete process described by Eq.~(\ref{process}). A standard method to periodically decouple the system and the environment is to instantaneously perform a nonselective projective $\si_z$ measurement on the qubit at the moments $n\tau$, $n\geq1$, thereby to project the qubit state onto the energy eigenstates, $|e\ra$ and $|g\ra$,
\begin{equation}\label{measurement}
\rho_S(n\tau)\mapsto \rho_S^M(n\tau)=\frac{1}{2}\left[\rho_S(n\tau)+\si_z\rho_S(n\tau)\si_z\right].
\end{equation}
Since the $\si_z$ projective measurement commutes with the bare Hamiltonian of the system, it serves as a quantum nondemolition (QND) measurement on the system. The effect of the QND measurement is retaining the qubit's $\si_z$-diagonal terms and erasing the off diagonals. It should be stressed that the measurements are nonselective, i.e., the measurement results are unread and not used to discard unwanted samples. A detailed dynamical description in the interval $[0, \tau]$ is provided in Appendix~\ref{PM}; undoubtedly it applies to any $[(n-1)\tau, n\tau], n=1,2,3,\cdots$.

Applying $n$ periodical measurements with a constant time spacing $\tau$, we obtain
\begin{equation}\label{rhoM}
\begin{aligned}
\rho_{ee}^M(t=n\tau)\approx&\left[\rho_{ee}(0)-\frac{J_+(\tau)}{J(\tau)}\right]
e^{-\frac{J(\tau)}{\tau}t}+\frac{J_+(\tau)}{J(\tau)},
\end{aligned}
\end{equation}
with $J(\tau)\equiv J_+(\tau)+J_-(\tau)$. Equation~(\ref{rhoM}) clearly presents a formal solution that, under the periodical measurements, the excitation population follows an exponential-like decay towards a quasisteady value $J_+(\tau)/J(\tau)$. The effective decay rate reads $J(\tau)/\tau$. Both steady state and decay rate rely on the measurement time interval $\tau$. In the Markovian limit with a sufficiently large $\tau\to\infty$, Eq.~(\ref{rhoM}) can reduce to
\begin{equation}\label{rMar}
\rho_{ee}^M(t)=\left[\rho_{ee}(0)-\rho_{ee}^B\right]e^{-\Ga_0 t}+\rho_{ee}^B,
\end{equation}
where $\Ga_0\equiv \lim_{\tau\to\infty}J_+(\tau)/\tau=2\pi[2n_T(\om_a)+1]G_0(\om_a)$ is the free decay rate as determined by Fermi's golden rule, and $\rho_{ee}^B\equiv \lim_{\tau\to\infty}J_+(\tau)/J(\tau)=(e^{\be\om_a}+1)^{-1}$, meaning that eventually the effective temperature of the system is equivalent to that of the thermal bath. It means that cooling cannot be realized in this limit.

\begin{figure}[htbp]\centering
\includegraphics[width=0.5\textwidth]{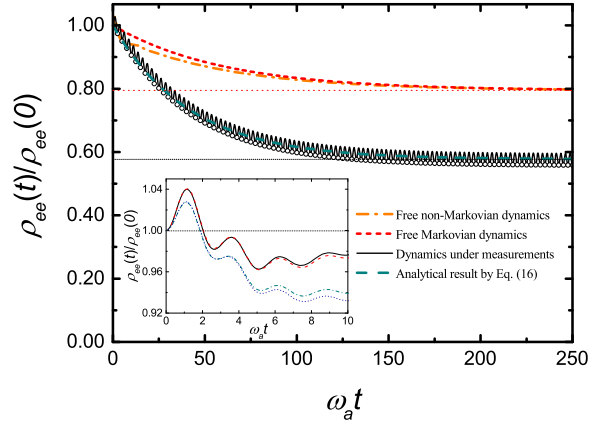}
\caption{(Color online) Main panel: excitation dynamics of the TLS in a finite-temperature bath. The initial population is $\rho_{ee}(0)=0.15$. The orange dot-dashed line represents the free evolution of the excitation population based on Eq.~(\ref{re}). The red short-dashed line represents the free dynamics in the Markovian limit based on Eq.~(\ref{rMar}). The black solid line represents excitation population over time under periodical measurements with a time interval $\om_a\tau=2.5$. The measurement moments are marked by the black circles. The green dashed line represents the exponential approximation of the black solid line based on Eq.~(\ref{rhoM}). The black short dotted and the red dotted horizontal lines depict the final excitation population with and without measurements, respectively. Inset: the normalized free dynamics of the excitation population under different initial conditions. The red dashed line and the black solid line are respectively the evolutions with and without the adiabatic approximation~(\ref{ree}) for initial population $\rho_{ee}(0)=0.12$, by which initially the effective temperature of the system is set to be equivalent to that of the bath. The blue short-dashed line and the green dot-dashed line are the evolutions with and without the adiabatic approximation for initial population $\rho_{ee}(0)=0.15$, respectively. Here the environmental spectrum is chosen as the Lorentzian type in Eq.~(\ref{Lorentz}), with $\al=0.01$, $\Lam=0.25\om_a$, and $\om_0=1.5\om_a$. The inverse temperature of the bath is set as $\be\om_a=2$.}
\label{rL}
\end{figure}

The excitation dynamics of the qubit in a finite-temperature bath, with and without measurements, is portrayed in the main panel of Fig.~\ref{rL}. Note, in this figure, the environmental temperature is fixed by $\be\om_a=2$ such that $\rho_{ee}(0)=0.12$ if the system is set to the same effective temperature. The red short-dashed line and the orange dot-dashed line reflect respectively the free normalized excitation population dynamics with and without the Markovian approximation. Both of them asymptotically decrease to the same steady-state value determined by the environmental temperature, although the system is started from an effective temperature higher than that of the environment due to $\rho_{ee}(0)=0.15$. Then we can see the effect of the periodical measurements on the excitation population by the numerical evaluation according to the time-convolutionless master equation~(\ref{re}) and the measurement projection~(\ref{measurement}) (see the black solid line). It can also be approximately captured by the analytical result~(\ref{rhoM}). Either decay rate of them is roughly larger than that in the free evolution and then approaches a new but lower steady value than that of the free evolution. To demonstrate that these results are insensitive to the initial condition and analytical technique, in the inset of Fig.~\ref{rL} we plot the free evolutions of the excitation population under two initial conditions. One of them is so selected that initially the separated system and environment are set to the same temperature and the other is selected as the same as that in the main panel. They are found to follow a quite similar dynamics: first rises a little bit in a very short time scale and then rapidly declines to a value lower than the initial one, so that it is always possible to have a proper time-spacing constant $\tau$ for nonselective measurements to reduce the excitation population as well as to cool down the system.

It is known that the effective decay rate is a crucial quantity to identify the quantum Zeno effect (QZE) and the quantum anti-Zeno effect (QAZE) in the open-quantum-system dynamics~(\ref{rMar}). In particular, the QZE occurs if the effective decay rate is smaller than the free decay rate, i.e., $J(\tau)/\tau<\Gamma_0$, and the QAZE does if $J(\tau)/\tau>\Gamma_0$. In this work, the effective decay rate $J(\tau)/\tau$ in Eq.~(\ref{rhoM}) is an extension to the results in Refs.~\cite{Kofman1996,Kofman2000,Zhang2018}, by including the contributions from the counter-rotating terms and replacing the SDF $G_0(\om)$ at zero temperature by $[2n_T(\om)+1]G_0(\om)$ at finite temperature. Accordingly, the sign of the second derivative of SDF $[2n_T(\om)+1]G_0(\om_a)$ can be regarded as a criterion~\cite{Zhang2018} to distinguish the QZE and QAZE.

The measurements also lead to a quasisteady state shared by the TLS and the environment. The effective temperature of the TLS could be different from that of the bath. When $T_S(\infty)$ at thermal equilibrium is higher than $T$, i.e., $J_+(\tau)/J(\tau)>(e^{\be\om_a}+1)^{-1}$, the TLS is heated up; otherwise, when $T_S(\infty)<T$, i.e., $J_+(\tau)/J(\tau)<(e^{\be\om_a}+1)^{-1}$, the TLS is cooled down.

A controversial problem emerging in previous works is whether or not the cooling of the system is equivalent to the QAZE. Based on the above analysis, the QAZE requires
\begin{equation}\label{QAZEc}
J(\tau)>\Ga_0\tau.
\end{equation}
whereas the cooling condition reads
\begin{equation}\label{Coolc}
J(\tau)>\Ga_0\frac{\int_0^\tau dt'\Ga_+(t')}{\Ga_+(\infty)},
\end{equation}
which can be obtained by Eqs.~(\ref{J}) and (\ref{rhoM}) with the relation $J_+(\infty)/J(\infty)=\Ga_+(\infty)/\Ga(\infty)=(e^{\be\om_a}+1)^{-1}$. It is clear that the right-hand side of Eqs.~(\ref{QAZEc}) and (\ref{Coolc}) are equivalent to each other in the Markovian limit $\Ga_+(t')=\Ga_+(\infty)$ or after a long time scale $\tau\to\infty$. Physically the QAZE is determined by the decay rate and the cooling relies on the thermal equilibrium state. Thus cooling is usually related to QAZE, but they should not be regarded as the same thing. In fact, a SDF with a large high-frequency profile has been found to be beneficial to the QAZE~\cite{Zhang2018} but not to the cooling, as the super-Ohmic model analyzed in the next section.

\section{Cooling conditions}\label{cool}

\subsection{General theory}\label{Gt}

Under periodical measurements, the TLS approaches a new quasisteady state given by
\begin{equation}
\rho_{ee}^M(\infty)=\rho_{ee}^B\left[1+M(\tau)\right].
\end{equation}
Here $M(\tau)$ is a dimensionless measurement-modified factor defined as
\begin{equation}\label{M}
M(\tau)=\frac{e^{\be\om_a}J_+(\tau)-J_-(\tau)}{J(\tau)}
=\frac{\int_0^\infty d\om F(\be,\be,\tau,\om)G_0(\om)}{J(\tau)},
\end{equation}
which represents the deviation from the new equilibrium established by measurements with time spacing $\tau$ to the old one without measurements. The filter function $F(\be,\be,\tau,\om)$ has been defined in Eqs.~(\ref{dr}) and (\ref{filter}). $M(\tau)<0$ and $M(\tau)>0$ can be respectively regarded as the cooling and the heating factors. It is independent on both the initial condition and the coupling strength in the weak-coupling regime (more details could be found in Appendix~\ref{appr}). The denominator of $M(\tau)$
\begin{equation}
\begin{aligned}
J(\tau)=&\tau^2\int_0^\infty d\om\left(\frac{2}{e^{\be\om}-1}+1\right)\\
&\times\left[{\rm sinc}^2\left(\frac{\om-\om_a}{2}\tau\right)+{\rm sinc}^2\left(\frac{\om+\om_a}{2}\tau\right)\right]G_0(\om)
\end{aligned}
\end{equation}
is always positive. Consequently, whether the system is cooled down or heated up is determined by the sign of the numerator and thus further determined by the sign of the filter function.

\begin{figure}[htbp]
\centering
\includegraphics[width=0.5\textwidth]{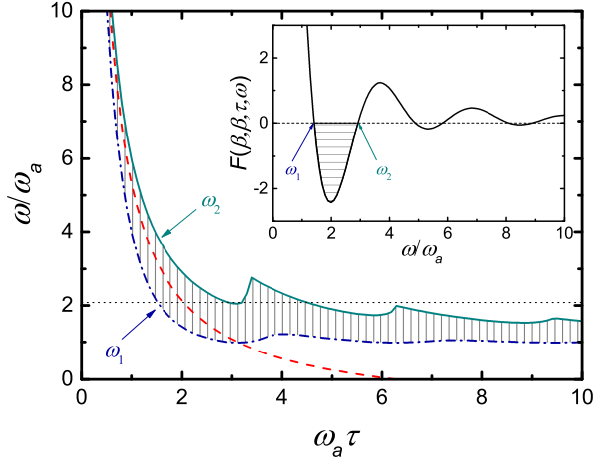}
\caption{(Color online) Main panel: the main frequency domain that could be used for cooling (the shaded area) with respect to the frequency of environmental mode vs the measurement time interval. Here the parameters are chosen as $\be_S(0)\om_a=\be\om_a=2$. The blue dot-dashed line and the green solid line are respectively the lower and upper bounds of the main effective domain for cooling. The red dashed line is a reference curve given by Eq.~(\ref{omtau}) as an analytical estimation of a frequency benchmark, around which the filter function $F^{cr}(\be,\be,\tau,\om)=0$ as a necessary cooling condition. The black dotted horizontal line depicts the measurement time-spacing-averaged $\om_2$ used to estimate the upper bound of cooling domain. Inset: an example of the filter function with $\om_a\tau=2$. The shaded area with negative values represents the main frequency-dependent cooling domain.}
\label{w12}
\end{figure}

The inset of Fig.~\ref{w12} plots an example of the filter function $F(\be,\be,\tau,\om)$ in Eq.~(\ref{M}) with an arbitrarily chosen environmental temperature and a measurement time interval. The frequencies $\om_1$ and $\om_2$ indicate the lower and upper bounds of the negative-value domain of the filter function, i.e., the main cooling domain, respectively. Throughout the whole range of frequency, the environmental-oscillators can be approximately divided into three parts: the oscillators with $\om<\om_1$ and $\om>\om_2$ contribute to heating the TLS; the oscillators with frequency $\om_1<\om<\om_2$ contribute to cooling the TLS. In the main panel of Fig.~\ref{w12}, we demonstrate the cooling domain by $\omega_1$ and $\omega_2$ as functions of the measurement time spacing $\tau$ (see the shaded area). When the measurement time spacing $\tau$ is comparatively small, both the lower bound $\om_1$ and upper bound $\om_2$ of the cooling domain remarkably decrease over $\tau$. Following the discussions in the end of Sec.~\ref{thermal}, the contribution from the counter-rotating filter function $F^{cr}$ defined in Eq.~(\ref{filter}) (in charge of heating) dominates that from the rotating-wave filter function $F^r$ (in charge of cooling) when $\tau$ is small and $\om>\om_a$. Thus the cooling domain with a short $\tau$ could be estimated by the condition in which the counter-rotating filter function $F^{cr}$ vanishes. Analytically it gives rise to
\begin{equation}\label{omtau}
\om=2\pi/\tau-\om_a
\end{equation}
as plotted as the red dashed line. As one can check in the main panel of Fig.~\ref{w12}, the cooling domain just appears around $2\pi/\tau-\om_a$ when $\omega_a\tau<3$ and $\om/\omega_a>1$. With a large measurement time spacing, the cooling domain is between $\om_a$ and the upper bound $\om_2$. The long measurement-time limit of $\om_2$ can be estimated by letting $F(\be,\be,\tau,\om)=0$, which resembles Eq.~(\ref{wup}) and yields
\begin{equation}\label{w2}
\om_2\approx \om_a[1+4\be\om_a\exp(-\be\om_a)],
\end{equation}
as plotted by the black-dotted line in the main panel. Accordingly, the upper bound of $\omega_2$ is about $2.47\omega_a$. Therefore, it is proper to use $\om=2\omega_a$ as a rough boundary to differentiate the contributions (heating or cooling) from various ranges of environmental spectrum.

It is experimentally challenging for many systems to enforce a number of measurements into an extremely small time scale. Physically, it is meaningful to consider the cooling condition for an appropriate $\tau$ such that the benchmark of cooling curve~(\ref{omtau}) actually stays in the near-resonant regime with respect to the system frequency $\om_a$. Thus one can divide the whole frequency domain into the low- and high-frequency domains and then obtain a more compact expression of the measurement-modified factor $M(\tau)$ with a not-too-small measurement time interval. Necessary details can be found in Appendix~\ref{appr}. In addition to the spectral analysis, we take the low-temperature limit, $e^{\be\om_a}\pm1\sim e^{\be\om_a}$ and $e^{-\be\om_a}\sim0$ as we focus on the cooling property of the bath. Eventually the measurement-modified factor $M(\tau)$~(\ref{M}) can be approximated as
\begin{widetext}
\begin{equation}\label{Mt}
M(\tau)\approx\frac{-\be[2G'_0(\om_a)-\be G_0(\om_a)]\left(\om_a-\frac{\sin\om_a\tau}{\tau}\right)
+\frac{\tau^2}{2}e^{\be\om_a}\int_0^{2\om_a} d\om {\rm sinc}^2\left(\frac{\om+\om_a}{2}\tau\right)G_0(\om)  +e^{\be\om_a}\int_{2\om_a}^\infty d\om\frac{G_0(\om)}{\om^2}} {G_0(\om_a)\left(\pi\tau-\frac{2}{\om_a}\right)
+\om_a G''_0(\om_a)+2\int_{2\om_a}^{\infty}d\om \frac{G_0(\om)}{\om^2}}.
\end{equation}
\end{widetext}
There are three terms in the numerator of $M(\tau)$~(\ref{Mt}). The first one represents the cooling effect from the oscillators which is near resonant with the TLS transition frequency. The second one is also from the near-resonant oscillators but stems from the counter-rotating terms, which oscillate rapidly compared to the first one. It determines a proper time interval $\tau$ for the minimum of temperature (cooling bound) achieved by measurements when the integral finds the minimum value. The third term describes the heating effect of the off-resonant oscillators. Its contribution becomes considerable when the SDF $G_0(\om)$ in the high-frequency regime grows faster than $\om$. The denominator of $M(\tau)$ increases with the time interval $\tau$, corresponding to the result that $M(\tau)\to 0$ with $\tau\to\infty$, i.e., without measurements.

To further simplify the result in Eq.~(\ref{Mt}), we assume the center of $G(\om)$ is not far off resonant from $\om_a$ and ignore the contribution from the oscillators with $\om>2\om_a$. Then in the large limit of $\tau$, we have
\begin{equation}\label{Mt1}
\begin{aligned}
M(\tau)\approx&-\be\om_a\frac{2G'_0(\om_a)-\be G_0(\om_a)} {G_0(\om_a)\left(\pi\tau-\frac{2}{\om_a}\right)
+\om_a G''_0(\om_a)}\\
&+\frac{e^{\be\om_a}\int_0^{2\om_a} d\om\frac{G_0(\om)}{(\om+\om_a)^2}} {G_0(\om_a)\left(\pi\tau-\frac{2}{\om_a}\right)+\om_a G''_0(\om_a)}.
\end{aligned}
\end{equation}
Equation~(\ref{Mt1}) loosely implies a necessary condition of the SDF $G_0(\om)$ to cool down the system: the logarithmic derivative of the SDF $G_0(\om)$ at $\om_a$ should be at least larger than $\be/2$, i.e., the cooling effect of the near-resonant oscillators should be dominative. We can deduce an analytical cooling criterion as
\begin{equation}\label{coolcondition}
\frac{G'_0(\om_a)}{G_0(\om_a)}>\frac{\be}{2}.
\end{equation}
Simultaneously the third term in the numerator of Eq.~(\ref{Mt}) indicates a second condition for cooling: the heating contribution from the off-resonant oscillators should be as small as possible, which means the SDF $G_0(\om)$ should have a sharp high-frequency cutoff and this cutoff frequency is not very far away from $\om_a$. In the following, we will check the above cooling factor $M(\tau)$ as well as these two conditions in two popular spectra: the modified Lorentzian model and the super-Ohmic model. They are found to be helpful to cool down the system in certain conditions.

\subsection{Modified Lorentzian model}

An environment often used in literature is described by the modified Lorentzian model with SDF~\cite{Erez2008,Gordon2009},
\begin{equation}\label{Lorentz}
G_0(\om)=\alpha \om\frac{\Lam^2}{\Lam^2+(\om-\om_0)^2},
\end{equation}
where $\al$ is a dimensionless coupling strength, $\om_0$ is the Lorentzian peak, and $\Lam$ is the Lorentzian width. This model can describe the environment of a cavity with not-so-high finesse mirrors, i.e., a leaky cavity.

\begin{figure}[htbp]
\centering
\includegraphics[width=0.5\textwidth]{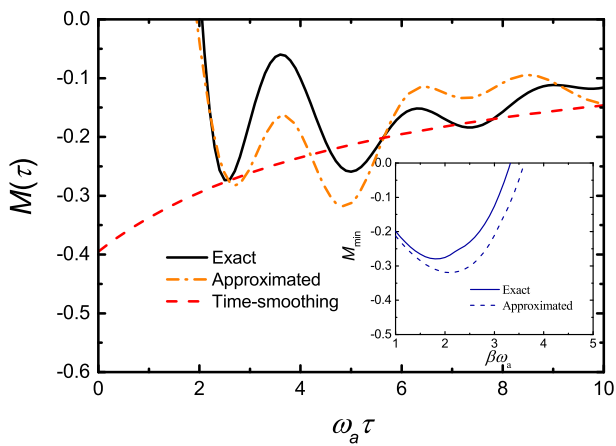}
\caption{(Color online) Main panel: the measurement-modified factor $M(\tau)$ vs the measurement interval $\om_a\tau$ for the modified Lorentzian model. Here $\al=0.01$, $\Lam=0.25\om_a$, and $\om_0=1.5\om_a$. The inverse temperature of the bath is set as $\be\om_a=2$. The black solid line, the orange dot-dashed line, and the red dashed line represent the exact~(\ref{M}), approximated~(\ref{Mt}) and time-smoothing~(\ref{Mt1}) results, respectively. Inset: the minimum $M_{\rm min}$ vs the inverse temperature $\be\om_a$. The solid and dashed lines represent the exact~(\ref{M}) and approximated~(\ref{Mt}) results, respectively. Other parameters are identical to those in the main panel.}
\label{ML}
\end{figure}

\begin{figure}[htbp]
\centering
\includegraphics[width=0.5\textwidth]{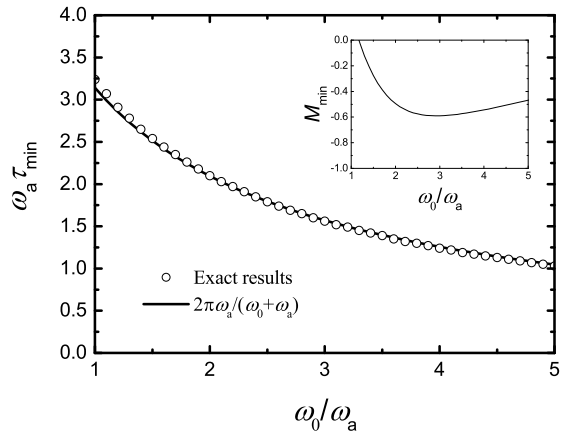}
\caption{Main panel: the optimal measurement time interval $\tau_{\rm min}$ for cooling vs the spectral peak $\om_0/\om_a$ in the modified Lorentzian model. The circles represent the exact results and the line is the reference line $2\pi\om_a/(\om_0+\om_a)$. Inset: the minimum $M_{\rm min}$ vs the peak $\om_0$. Parameters are chosen as the same as those of Fig.~\ref{ML}.}
\label{Lw0}
\end{figure}

The numerical results of the measurement-modified factor $M(\tau)$ for this model are demonstrated by the black solid line, the orange dot-dashed line and the red dashed line in the main panel of Fig.~\ref{ML}. These results for the same parameters of the modified Lorentzian model are obtained respectively by the exact~(\ref{M}), the approximated~(\ref{Mt}) with respect to large $\tau$ and low temperature, and the time-smoothing~(\ref{Mt1}) expression of the cooling factor $M(\tau)$. Note we only present the cooling results with $M(\tau)\leq0$, which occurs with a large time spacing about $\tau\gtrsim2/\omega_a$ and fluctuates over $\tau$. It is shown that our approximated analytical result (see the orange dot-dashed line) could catch the main features of $M(\tau)$ and also mimic the following fluctuations, while the time-smoothing result from Eq.~(\ref{Mt1}) shown by the red dashed line can outline the main tendency of the exact result in the regime of large $\tau$.

Both the black solid line and the orange dot-dashed line indicate that an optimized measurement time spacing $\tau$ could be numerically obtained: a minimum value of $M_{\rm min}(\tau)$ corresponds to a maximal cooling efficiency. In the inset of Fig.~\ref{ML}, we depict the minimum measurement-modified factor as a function of the inverse temperature $\be$ of the bath. From Eq.~(\ref{Mt1}), the absolute value of both the cooling effect [see the first line of Eq.~(\ref{Mt1})] and the heating effect [see the second line of Eq.~(\ref{Mt1})] increase with an increasing inverse temperature $\beta$. Then the competition between them leads to a nonmonotonic behavior of the minimum measurement-modified factor $M_{\rm min}$ with respect to $\be$, so that the cooling efficiency could be also optimized with a moderate environmental temperature. Our approximation from Eq.~(\ref{Mt}) given by the dashed line fits well with the exact result given by the solid line.

The modified Lorentzian model with a distinguished peak and comparatively low wings is a good testbed to check the cooling conditions for the environment that we have proposed by the general theory in Sec.~\ref{Gt}. One can choose a proper measurement time interval to locate the spectral peak to the cooling domain and then efficiently cool down the open system. We plot the optimal measurement time interval $\tau_{\rm min}$ for cooling as a function of the peak $\om_0$ (the circles) in the main panel of Fig.~\ref{Lw0}. According to Eq.~(\ref{omtau}), when the spectral peak $\om_0$ is off resonant with respect to the system frequency, the maximal cooling is attained nearly at $\tau\approx2\pi/(\om_0+\om_a)$, where the heating contribution from the counter-rotating terms $F^{cr}(\be,\be,\tau,\om)$ vanishes. To achieve the lowest temperature, one has to enforce more frequent measurements into the dynamics in the case of a larger spectral peak. As for the particular minimum measurement-modified factor $M_{\rm min}$, the results shown in the inset of Fig.~\ref{Lw0} indicate that it is also optimized with a moderate peak $\om_0$.

\subsection{Super-Ohmic model}

Another widely used environment in the thermodynamics and solid-state physics is the super-Ohmic model. The SDF of this model reads~\cite{Leggett1987}
\begin{equation}
G_0(\om)=\al\om_c^{1-s}\om^s\Theta(1-\om/\om_c).
\end{equation}
Here $\al$ is a dimensionless coupling parameter, $\Theta(1-\om/\om_c)$ is a sharp cutoff function, and $\om_c$ is the cutoff frequency. A particular example of this general model is the Debye model, which has a well-known expression $G_0(\om)=\al\om_c^{-2}\om^3\Theta(1-\om/\om_c)$. The cubic frequency dependence of the Debye model arises from the coupling strength $g_k\propto\sqrt{\om_k}$ and the Debye density of states $\sum_k\de(\om-\om_k)\propto \om^2$. The cutoff frequency $\om_c$ is the Debye frequency in this example.

\begin{figure}[htbp]
\centering
\includegraphics[width=0.5\textwidth]{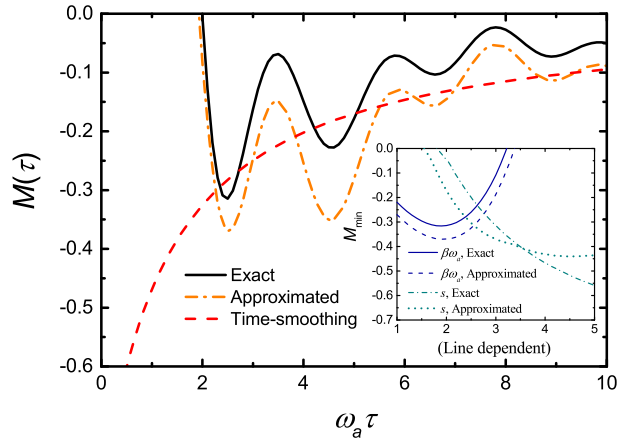}
\caption{(Color online) Main panel: the measurement-modified factor $M(\tau)$ vs the time interval $\om_a\tau$ for the Debye model. Parameters are chosen as $\al=0.01$ and $\om_c=2\om_a$. The inverse temperature of the bath is set as $\be\om_a=2$. The black solid line, the orange dot-dashed line and the red dashed line represent the exact~(\ref{M}), approximated~(\ref{Mt}), and time-smoothing~(\ref{Mt1}) results, respectively. Inset: the minimum $M_{\rm min}$ vs the inverse temperature $\be\om_a$ (blue) and the exponent $s$ (green). The solid and dashed lines represent the exact~(\ref{M}) and approximated~(\ref{Mt}) results for different $\be\om_a$, respectively. The dot-dashed and short-dashed lines represent the exact~(\ref{M}) and approximated~(\ref{Mt}) results with different $s$, respectively.}
\label{MD}
\end{figure}

The measurement-modified factor $M(\tau)$ for the Debye model versus the measurement time spacing $\tau$ is demonstrated by the three lines in the main panel of Fig.~\ref{MD}. Again it shows that our analytical approximations in Eqs.~(\ref{Mt}) and (\ref{Mt1}) could be used to quantitatively describe the exact result of $M(\tau)$. Roughly a moderate $\tau$ gives rise to an efficient cooling effect implied by a negative $M(\tau)$ and the absolute value of $M(\tau)$ declines with increasing $\tau$. Similar to the Lorentzian model, the minimum measurement-modified factor $M_{\rm min}$ behaves nonmonotonically with respect to the inverse temperature of the bath (see the inset of Fig.~\ref{MD}). Furthermore, we plot the minimum measurement-modified factor $M_{\rm min}$ versus the bath exponent $s$ by the green dot-dashed line in the inset of Fig.~\ref{MD}. We can find that $|M_{\rm min}|$ is enhanced with increasing $s$, while the cooling is conditioned by $s\geq2$. Our approximation given by Eq.~(\ref{Mt}) (see the short-dashed-line) fits well with the exact result. In fact one can deduce from Eqs.~(\ref{Mt1}) and (\ref{coolcondition}) that the exponent $s$ for the super-Ohmic model dramatically affects the cooling effect. From the loose cooling condition~(\ref{coolcondition}), it could be found that the maximum inverse temperature (minimum temperature) $\be_{\rm max}=2s/\om_a$, so that, as described by the green lines in the inset of Fig.~\ref{MD}, a larger $s$ gives rise to a lower temperature that the TLS could reach by measurements.

\section{Conclusion}\label{conc}

In summary, we have explicitly investigated the required spectral structure of the bath to cool down an open quantum system coupled to a finite-temperature environment in the context of nonselective and nondemolition measurements. The memory effect of the bath that yields time-dependent damping rates is found to be the primary prerequisite of cooling. One can exploit the modified dynamics of the open system as well as a time-dependent energy flow from the system to the bath. It might be reminiscent of the non-Markovian effect of the environment. Yet a scrutinization over the relationship between any recent result including cooling and the non-Markovianity~\cite{Hall2014,Li2018} could be performed in a future work. The secondary prerequisite is performing the nonselective measurements to periodically decouple the system and the environment, which can also be replaced by alternative ways of measurements or control. To obtain the cooling conditions particularly relevant to the environment structure, we have derived a compact form of a time-convolutionless master equation for the system under a finite-temperature bath.

Our model is a standard spin-boson model without rotating-wave approximation. We find that the contribution from the counter-rotating terms cannot be neglected especially at short time scales and low temperature. Considering both the rotating-wave and the counter-rotating terms, the environmental bosonic oscillators within the near-resonant regime with respect to the system transition frequency $\om_a$ can be used to cool down the system, while the oscillators with lower or higher frequencies can heat the system up. Thus to realize the cooling of the system, one should strengthen coupling between the system and the near-resonant modes and meanwhile weaken the influence from the far-off-resonant modes. Based on this idea, we propose two cooling conditions with regard to the bath structure: (i) the logarithmic derivative of the spectrum around $\om_a$ should be as large as possible; (ii) the spectrum should have a sharp high-frequency cutoff. Based on these conditions we deduced, one can roughly estimate whether or not an environment could allow to cool down the system. Both of them should be simultaneously satisfied. For example, the $1/f$-noise model is popular in solid-state environments, which ranges from the voltages and currents in vacuum tubes to the diodes and transistors~\cite{Niemann2013}. The negative logarithmic derivative of its SDF throughout the whole frequency domain makes it impossible to cool down the quantum systems, although it satisfies the condition (ii). We have checked our conditions in the Lorentzian model and the Debye model, and thus searched the proper parameters for cooling. It is found that the environment with SDF in the Lorentzian form or the super-Ohmic form can be efficiently used to cool the system by discrete measurements.

\section*{Acknowledgments}

We acknowledge grant support from the National Science Foundation of China (Grants No. 11575071 and No. U1801661), Zhejiang Provincial Natural Science Foundation of China under Grant No. LD18A040001, and the Fundamental Research Funds for the Central Universities.

\appendix

\section{Second-order time-convolutionless master equation}\label{SNME}

Consider a general total Hamiltonian of the system and the environment:
\begin{equation}
H=H_S+H_B+H_I(t)
\end{equation}
where $H_S$ and $H_B$ are the Hamiltonians for the system and the environment, respectively, and $H_I(t)$ is the interaction between them. It is convenient to go to the interaction picture in which
\begin{equation}
\label{LE}
\frac{d}{dt}\ti{\rho}(t)=-i[\ti{H}_I(t),\ti{\rho}(t)],
\end{equation}
where $\ti{H}_I(t)=e^{iH_0t}H_I(t)e^{-iH_0t}$ and $\ti{\rho}(t)=e^{iH_0t}\rho(t)e^{-iH_0t}$ is the combined density-matrix operator at the moment $t$ in the interaction picture with $H_0=H_S+H_B$. One can formally integrate Eq.~(\ref{LE}) to obtain
\begin{equation}
\ti{\rho}(t)=\ti{\rho}(0)-i\int_0^t dt'\left[\ti{H}_I(t'),\ti{\rho}(t')\right]
\end{equation}
and then substitute it into the commutator in Eq.~(\ref{LE}). The resultant equation is
\begin{equation}\label{LE2}
\frac{d}{dt}\ti{\rho}(t)=-i\left[\ti{H}_I(t),\ti{\rho}(0)\right]-\int_0^t dt'\left[\ti{H}_I(t),\left[\ti{H}_I(t'),\ti{\rho}(t')\right]\right].
\end{equation}
Suppose that initially $\rho(0)=\ti{\rho}(0)=\rho_S(0)\otimes\rho_B$, where $\rho_B$ is a stationary state of the environment, i.e., the system and the environment are initially uncorrelated. Then, after tracing over the degrees of freedom of the environment, Eq.~(\ref{LE2}) gives the exact master equation of the reduced density matrix in the interaction picture:
\begin{equation}
\frac{d}{dt}\ti{\rho}_S(t)=-\int_0^t dt'{\rm Tr}_B\left\{\left[\ti{H}_I(t),\left[\ti{H}_I(t'),\ti{\rho}(t')\right]\right]\right\},
\end{equation}
where it is assumed that
\begin{equation}\label{assu}
{\rm Tr}_B\left[\ti{H}_I(t)\rho_B\right]=0,
\end{equation}
to eliminate the first term in Eq.~(\ref{LE2}). A popular example that satisfies Eq.~(\ref{assu}) is that the environment is coupled with the system by linear operators and it is in a thermal equilibrium state or any state in the form $\rho_B=\sum_nc_n|n\ra\la n|$, where $|n\ra$ is a product of Fock states for all the modes.

The environment contains a sufficiently large number of modes so that the backaction of the system on the environment is ignorable in the weak-coupling regime. The exact full state $\rho(t)$ is supposed to close to $\rho_S(t)\otimes\rho_B$, which is called the \emph{Born approximation}. One then obtains
\begin{equation}
\frac{d}{dt}\ti{\rho}_S(t)=-\int_0^t dt'{\rm Tr}_B\left\{\left[\ti{H}_I(t),\left[\ti{H}_I(t'),\ti{\rho}_S(t')\otimes\rho_B\right]\right]\right\}.
\end{equation}
We often assume the interaction Hamiltonian is
\begin{equation}
H_I(t)=\sum_j\left[L_j^\dag B_j(t)+L_j B_j^\dag(t)\right],
\end{equation}
where $L$ and $B$ are operators of the system and the bath, respectively. We eventually obtain the master equations in the Schr\"{o}dinger picture,
\begin{equation}\label{ME1}
\begin{aligned}
&\frac{d}{dt}\rho_S(t)=-i[H_S,\rho_S(t)]+\sum_{jj'}\int_0^t dt'\Big\{ \\
&\left[e^{-iH_S(t-t')}L_{j'}^\dag e^{iH_S(t-t')}\rho_S(t'),L_j^\dag\right]\Phi_{B_jB_{j'}}(t-t')\\
+&\left[e^{-iH_S(t-t')}L_{j'} e^{iH_S(t-t')}\rho_S(t'),L_j^\dag\right]\Phi_{B_j B_{j'}^\dag}(t-t') \\
+&\left[e^{-iH_S(t-t')}L_{j'}^\dag e^{iH_S(t-t')}\rho_S(t'),L_j\right]\Phi_{B_j^\dag B_{j'}}(t-t')\\
+&\left[e^{-iH_S(t-t')}L_{j'} e^{iH_S(t-t')}\rho_S(t'),L_j\right]\Phi_{B_j^\dag B_{j'}^\dag}(t-t')\\ +&{\rm H.c.}\Big\}
\end{aligned}
\end{equation}
where we have defined the correlation functions of the bath as
\begin{equation}
\Phi_{XY}(t-t')={\rm Tr}_B\left[\ti{X}(t)\ti{Y}(t')\rho_B\right]
\end{equation}
with $\ti{A}(t)\equiv e^{iH_Bt}A(t)e^{-iH_Bt}$ and $A=X,Y$.

Next we consider a quantum two-level system coupled to a bosonic environment with a general time-dependent Hamiltonian of the form
\begin{equation}
\label{AH}
\begin{gathered}
H_S=\frac{1}{2}\om_a\si_z,\\
H_B=\sum_k\om_k a_k^\dag a_k,\\
L_1=\si_-,\quad L_2=\si_+,\\
B_1=B_2=B=\ep(t)\sum_k g_k a_k.\\
\end{gathered}
\end{equation}
The bath correlation functions are then
\begin{equation}
\label{ACF}
\begin{aligned}
\Phi_{BB}(t-t')=&0,\\
\Phi_{BB^\dag}(t-t')=&\ep(t)\ep^*(t')\sum_k |g_k|^2[1+n_T(\om_k)]e^{-i\om_k (t-t')},\\
\Phi_{B^\dag B}(t-t')=&\ep^*(t)\ep(t')\sum_k |g_k|^2n_T(\om_k)e^{i\om_k (t-t')},\\
\Phi_{B^\dag B^\dag}(t-t')=&0,
\end{aligned}
\end{equation}
where $n_T(\om)=(e^{\be\om}-1)^{-1}$ is the average population of the oscillator (bath-mode) with frequency $\om$. Then we make a Markovian approximation by replacing $\rho_S(t')$ with $\rho_S(t)$ in Eq.~(\ref{ME1}) and decompose the integrations for the coefficients in Eq.~(\ref{ME1}) into the real and imaginary parts:
\begin{equation}
\label{AGD}
\begin{gathered}
\int_0^t dt' \left[e^{i\om_a(t-t')}\Phi_{BB^\dag}(t-t')\right]
=\frac{\Ga_-^r(t)}{2}+i\De_-^r(t),\\
\int_0^t dt'\left[e^{-i\om_a(t-t')}\Phi_{B^\dag B}(t-t')\right]
=\frac{\Ga_+^r(t)}{2}+i\De_+^r(t),\\
\int_0^t dt'\left[e^{-i\om_a(t-t')}\Phi_{BB^\dag}(t-t')\right]
=\frac{\Ga_+^{cr}(t)}{2}+i\De_+^{cr}(t),\\
\int_0^t dt'\left[e^{i\om_a(t-t')}\Phi_{B^\dag B}(t-t')\right]
=\frac{\Ga_-^{cr}(t)}{2}+i\De_-^{cr}(t).
\end{gathered}
\end{equation}
Eventually we obtain a time-convolutionless master equation:
\begin{equation}
\label{AME}
\begin{aligned}
\frac{d}{dt}\rho_S(t)=&-i\left[H_S+\sum_{j=\pm}\De_j(t)\si_j^\dag \si_j,\rho_S(t)\right]\\
&+\sum_{j=\pm}\frac{\Ga_j(t)}{2}\mL[\si_j](\rho_S(t))\\
&+\sum_{j=\pm}\left\{\left[\frac{\Ga_j(t)}{2}+i\De_j(t)\right]\si_j\rho_S(t)\si_j+{\rm H.c.}\right\},
\end{aligned}
\end{equation}
where $\De_j(t)=\De_j^r(t)+\De_j^{cr}(t)$, $\Ga_j(t)=\Ga_j^r(t)+\Ga_j^{cr}(t)$, and the Lindblad superoperator is defined as
\begin{equation}
\mL[\si](\rho)\equiv 2\si\rho\si^\dag-\{\si^\dag \si,\rho\}.
\end{equation}
Previously, this master equation has been obtained by the time-convolutionless projection technique~\cite{Breuer1999,Breuer2002,Kofman2004,Zeng2012}. In general, the Markovian approximation appears as an additional approximation after the Born approximation. Both approximations are only valid to the second order with respect to the coupling strength.

It should be noted that Eq.~(\ref{AME}) applied to the time-dependent interaction Hamiltonian in Eq.~(\ref{AH}), while, in the main text of this work, we focus on the time-independent interaction Hamiltonian as indicated by Eq.~(\ref{H}). Therefore, here we merely let $\ep(t)=1$ when using Eq.~(\ref{AME}) or Eq.~(\ref{ME}). Also the Lamb shifts and transition rates in Eqs.~(\ref{De}) and (\ref{Ga}) are obtained by substituting $\ep(t)=1$ into Eqs.~(\ref{ACF}) and (\ref{AGD}).

\section{Decoupling the system and the environment by non-demolition measurement }\label{PM}

Suppose we attach the open TLS system with a detector modeled as another qubit to perform nondemolition measurement. The interaction Hamiltonian between the system and the detector is assumed to be:
\begin{equation}
H_{SD}(t)=h(t)|e\ra\la e|(|0\ra\la 0|+|1\ra\la 1|-|0\ra\la 1|-|1\ra\la 0|),
\end{equation}
where $|0\ra$ and $|1\ra$ are two energy-degenerate states of the detector qubit and
\begin{equation}
h(t)=\frac{\pi}{2\tanh(1)\tau_D}\left[\tanh^2\left(\frac{t}{\tau_D}\right)-1\right]
\end{equation}
is a smooth temporal profile of the system coupling to the detector qubit during the measurement time $\tau_D$. Taking the measurement intervals to be $[0, \tau_D]$, the time-evolution operator $U_D=\exp[-i\int_0^{\tau_D}dt'H_{SD}(t')]$ would serve as a controlled-NOT (CNOT) operation. In other words,
\begin{equation}
\begin{aligned}
U_D|g\rangle|0\rangle=&|g\rangle|0\rangle, \\
U_D|g\rangle|1\rangle=&|g\rangle|1\rangle, \\
U_D|e\rangle|0\rangle=&|e\rangle|1\rangle, \\
U_D|e\rangle|1\rangle=&|e\rangle|0\rangle.
\end{aligned}
\end{equation}
Note here $\tau_D$ is so short compared to all the other timescales (the impulsive limit) that the free evolution of the system under the environment is switched off during the period of measurement.

Preparing the initial state of the detector to be $|0\ra$ and assuming the state of the detector is unread after measurement (by tracing out the degrees of freedom of the detector),
\begin{equation}
\begin{aligned}
\rho_S\mapsto&{\rm Tr}_D\left(U_D\rho_S\otimes|0\ra\la 0|U_D^\dag\right)\\
=& |e\ra\la e|\rho_S|e\ra\la e|+|g\ra\la g|\rho_S|g\ra\la g|,
\end{aligned}
\end{equation}
where the population of the TLS are preserved and the off-diagonal elements are erased. Though the system is entangled with the bath during the free evolution, the combined density matrix of the system and the environment $\rho$ after measurement becomes
\begin{equation}
\rho\mapsto{\rm Tr}_D (U_D\rho\otimes|0\ra\la 0|U_D^\dag)= B_{ee}|e\ra\la e|+B_{gg}|g\ra\la g|,
\end{equation}
where $B_{ee(gg)}=\la e(g)|\rho_S|e(g)\rangle\rho_B$ is the environmental state correlated to $|e(g)\ra$. It is shown that the postmeasurement state $\rho$ is approximately equivalent to a product state of the TLS and the bath under Born approximation~\cite{Erez2008,Gordon2009}, so that this postmeasurement technique actually provides a roadway to the Nakajima-Zwanzig superprojection~\cite{Breuer2002}.

\section{Approximation of the measurement-modified factor $M(\tau)$}\label{appr}

In this work, the frequency or energy $2\om_a$ is found to be a proper boundary between the low-frequency (long wavelength) and the high-frequency (short wavelength) regimes; the numerator of $M(\tau)$ defined in Eq.~(\ref{M}) can be decomposed into the following four terms:
\begin{equation}\label{nume}
\begin{aligned}
&e^{\be\om_a}J_+(\tau)-J_-(\tau)=\\
-&\tau^2\int_0^{2\om_a} d\om\frac{e^{\be\om}-e^{\be\om_a}}{e^{\be\om}-1}{\rm sinc}^2\left(\frac{\om-\om_a}{2}\tau\right)G_0(\om)\\
-&\tau^2\int_{2\om_a}^\infty d\om\frac{e^{\be\om}-e^{\be\om_a}}{e^{\be\om}-1}{\rm sinc}^2\left(\frac{\om-\om_a}{2}\tau\right)G_0(\om)\\
+&\tau^2\int_0^{2\om_a} d\om \frac{e^{\be\om}-e^{-\be\om_a}}{e^{\be\om}-1}e^{\be\om_a}{\rm sinc}^2\left(\frac{\om+\om_a}{2}\tau\right)G_0(\om)\\
+&\tau^2\int_{2\om_a}^\infty d\om\frac{e^{\be\om}-e^{-\be\om_a}}{e^{\be\om}-1}e^{\be\om_a}{\rm sinc}^2\left(\frac{\om+\om_a}{2}\tau\right)G_0(\om).
\end{aligned}
\end{equation}
The function ${\rm sinc}^2\left[(\om-\om_a)\tau/2\right]$ in the first term implies that the energy exchange between the system and the near-resonant oscillators is much more efficient than that between the system and the detuning bath oscillators. Thus we expand $\frac{e^{\be\om}-e^{\be\om_a}}{e^{\be\om}-1}G_0(\om)$ by the Taylor series to the second order with respect to $\omega_a$ and then the first term in Eq.~(\ref{nume}) can be approximated as
\begin{equation}\label{first}
\begin{aligned}
&-\tau^2\int_0^{2\om_a} d\om\frac{e^{\be\om}-e^{\be\om_a}}{e^{\be\om}-1}{\rm sinc}^2\left(\frac{\om-\om_a}{2}\tau\right)G_0(\om)\\
\approx &-2\int_0^{2\om_a}d\om\sum_{n=0}^2\left.\left[\frac{e^{\be\om}-e^{\be\om_a}}{e^{\be\om}-1}G_0(\om)\right]^{(n)}\right|_{\om=\om_a}\\ &\times\frac{[1-\cos(\om-\om_a)\tau](\om-\om_a)^{n-2}}{n!}\\
=&-2\left.\left[\frac{e^{\be\om}-e^{\be\om_a}}{e^{\be\om}-1}G_0(\om)\right]^{(2)}\right|_{\om=\om_a}
\left(\om_a-\frac{\sin\om_a\tau}{\tau}\right)\\
=&-\frac{2\be e^{\be\om_a} [2(e^{\be\om_a}-1)G'_0(\om_a)-\be(e^{\be\om_a}+1)G_0(\om_a)]}{(e^{\be\om_a}-1)^2}\\
&\times\left(\om_a-\frac{\sin\om_a\tau}{\tau}\right)\\
\approx &-2\be[2G'_0(\om_a)-\be G_0(\om_a)]
\left(\om_a-\frac{\sin\om_a\tau}{\tau}\right),
\end{aligned}
\end{equation}
where we have used the low-temperature limit $e^{\be\om_a}\pm1\sim e^{\be\om_a}$ in the last line.

In the high-frequency regime, the value of the integration is insensitive to the rapidly oscillating behavior of the filter function, and one can then simply replace the square sine function by its mean value $1/2$
\begin{equation}
\tau^2{\rm sinc}^2\left(\frac{\om\pm\om_a}{2}\tau\right)\approx\tau^2\frac{1/2}{[(\om\pm\om_a)\tau/2]^2}
=\frac{2}{(\om\pm\om_a)^2},
\end{equation}
so that the second term in Eq.~(\ref{nume}) is approximated by
\begin{equation}
\begin{aligned}
&-\tau^2\int_{2\om_a}^\infty d\om\frac{e^{\be\om}-e^{\be\om_a}}{e^{\be\om}-1}{\rm sinc}^2\left(\frac{\om-\om_a}{2}\tau\right)G_0(\om)\\
\approx &-2\int_{2\om_a}^\infty d\om\frac{e^{\be\om}-e^{\be\om_a}}{e^{\be\om}-1}\frac{G_0(\om)}{(\om-\om_a)^2}\\
\approx &-2\int_{2\om_a}^\infty d\om\frac{G_0(\om)}{\om^2},
\end{aligned}
\end{equation}
Similarly, the third and fourth terms in Eq.~(\ref{nume}) and the denominator of $M(\tau)$ can be approximated as
\begin{equation}
\begin{aligned}
&\tau^2\int_0^{2\om_a} d\om \frac{e^{\be\om}-e^{-\be\om_a}}{e^{\be\om}-1}e^{\be\om_a}{\rm sinc}^2\left(\frac{\om+\om_a}{2}\tau\right)G_0(\om)\\
\approx &\tau^2e^{\be\om_a}\int_0^{2\om_a} d\om {\rm sinc}^2\left(\frac{\om+\om_a}{2}\tau\right)G_0(\om),
\end{aligned}
\end{equation}
\begin{equation}
\begin{aligned}
&\tau^2\int_{2\om_a}^\infty d\om\frac{e^{\be\om}-e^{-\be\om_a}}{e^{\be\om}-1}e^{\be\om_a}{\rm sinc}^2\left(\frac{\om+\om_a}{2}\tau\right)G_0(\om)\\
\approx &2\int_{2\om_a}^\infty d\om\frac{e^{\be\om}-e^{-\be\om_a}}{e^{\be\om}-1}e^{\be\om_a}\frac{G_0(\om)}{(\om+\om_a)^2}\\
\approx &2 e^{\be\om_a}\int_{2\om_a}^\infty d\om\frac{G_0(\om)}{\om^2},
\end{aligned}
\end{equation}
and
\begin{equation}
J(\tau)\approx2G_0(\om_a)(\pi\tau-\frac{2}{\om_a})
+2\om_a G''_0(\om_a)+4\int_{2\om_a}^{\infty}d\om \frac{G_0(\om)}{\om^2},
\end{equation}
respectively.

Eventually, we obtain an approximated expression of the measurement-modified factor $M(\tau)$ in Eq.~(\ref{Mt}),
\begin{widetext}
\begin{equation}
M(\tau)\approx\frac{-\be[2G'_0(\om_a)-\be G_0(\om_a)]
\left(\om_a-\frac{\sin\om_a\tau}{\tau}\right)
+\frac{\tau^2}{2}e^{\be\om_a}\int_0^{2\om_a} d\om {\rm sinc}^2\left(\frac{\om+\om_a}{2}\tau\right)G_0(\om)  +e^{\be\om_a}\int_{2\om_a}^\infty d\om\frac{G_0(\om)}{\om^2}} {G_0(\om_a)\left(\pi\tau-\frac{2}{\om_a}\right)
+\om_a G''_0(\om_a)
+2\int_{2\om_a}^{\infty}d\om \frac{G_0(\om)}{\om^2}}.
\end{equation}
\end{widetext}

\bibliographystyle{apsrevlong}
\bibliography{Cooling}

\end{document}